\documentstyle[11pt]{article}
\title{The Soret and Dufour Effects in Statistical Dynamics}
\author{R. F. Streater,\\Dept. of Maths.,\\ King's College,\\ Strand,\\
London, WC2R 2LS}
\date{10/6/1998}
\begin{document}
\maketitle
\begin{abstract}
We set up a discrete space-time dynamical model of molecules
with thermalised kinetic energy and repulsive cores, in an external
potential. The model obeys the first and second laws of thermodynamics.
The continuum limit, obtained using a MAPLE program, gives rise to
coupled reaction-diffusion equations for the density and temperature fields.
The system obeys Onsager symmetry and exhibits the Soret and Dufour effects.
\end{abstract}
\section{Introduction}
It was known from experiments in the Nineteenth Century that a liquid
at uniform temperature, but with a gradient in its concentration, develops
a non-uniform temperature (the Dufour effect); this was confirmed by
Waldmann \cite{Waldmann}. Thus Fourier's law \cite{Fourier} needs
modification. A local version of the Thompson effect \cite{Thompson}
was also found, in 1856, by Ludwig: if the density of a liquid is
uniform, but the temperature is nonuniform, then there is a transport of matter.
Thus, Fick's law \cite{Fick} needs modification. The
ratio of the molecular migration to the temperature gradient is called the
Soret coefficient, after C. Soret, who studied the phenomenon in
1879-81 \cite{Soret}. For gases, the
Soret effect, under the name `thermal diffusion', was obtained
theoretically in 1911 by Enskog \cite{Enskog} using kinetic theory, and
also by Chapman in 1912 \cite{Chapman1}. In Enskog's work, it showed up
in the Lorentzian gas but not in the Maxwellian gas. It was first observed
experimentally in gas mixtures by Chapman and Dootson \cite{Dootson}.

Concerning Enskog's work, Hirschfelder et al. say ``Each time we lower
the level of description it is necessary to introduce a condition which
restricts the possible states under consideration. In this case [Enskog's
method] it is not clear how the restriction has been imposed'' 
\cite{Hirschfelder}, p. 492. In the present paper we derive a model of a
dense fluid using the methods of statistical dynamics \cite{RFS1}, which
is a systematic way to reduce the level of description. The possible states
are restricted using information geometry in a well-defined way 
\cite{Ingarden,Balian}. In the present model, the essential part of
the dynamics, the random
hopping of molecules to neighbouring holes, is included, but there is no
interparticle potential. The kinetic energy of the particles is fully
thermalised, in the spirit of Smoluchowski \cite{Smol}. We find that the
model exhibits both the Dufour and the Soret effects, while obeying the
first and second laws of thermodynamics. This success seems to contradict
the statement of \cite{Chapman2}, p. 103: ``No really satisfactory simple
theory of this thermal diffusion can be given...The reason is that thermal
diffusion is an interaction phenomenon. Similar remarks apply to the
inverse `diffusion thermo-effect' [the Dufour effect in gases]''. Our
theory, statistical dynamics, is not as simple as that using free paths
referred to in this quotation, and it may be a matter of opinion whether
it is `really satisfactory'; however we do show that the effects follow
from the assumptions that the state is in local thermal equilibrium, and
that the hopping rate is proportional to the kinetic energy. These are
kinematic assumptions, and do not require the solving of a model with an
explicit interaction between the particles. The interaction enters only
implicitly; its effect is replaced by the hopping term and the exclusion
principle, followed by local thermalisation.
So the last part of the quotation is not true.

In our model, the system is described by the particle density $\rho(x,t)$
and the temperature field $\Theta(x,t)$. The potential energy of a particle
at $x$ is $V(x)$, and the heat capacity is unity. Thus the density of heat
is $\rho\Theta$. There is a maximum possible density, denoted $\rho_m$;
this corresponds to a hard core of diameter $\ell$, where $\rho_m=
\ell^{-\nu}$ in $\nu$ dimensions. The particle current $j_c$, and the heat
current $j_\gamma$, are given in terms of $\rho,\;\Theta$ by
\begin{eqnarray}
j_c&=&-\lambda\left(\Theta\nabla\rho+\rho(1-\rho/\rho_m)\nabla(\Theta+V)
\right)\label{eq-ncurrent}\\
j_\gamma&=&2\left(\Theta j_c-\lambda \rho(1-\rho/\rho_m)\Theta\nabla\Theta
\right).
\label{eq-continuum}
\end{eqnarray}
Here $\lambda$ is the microscopic hopping rate. These relations in
conjunction with the conservation laws
\begin{equation}
\frac{\partial\rho}{\partial t}+\mbox{div\,}j_c=0;
\hspace{.6in}\frac{\partial(\rho\Theta)}
{\partial t}+\mbox{div\,}j_\gamma=-j_c.\nabla V
\label{eq-conserve}
\end{equation}
determine the dynamics. The heat current $j_\gamma$ is not conserved,
because of the heat source $-j_c.\nabla V=j_c.F$, where $F$ is the force.
Thus the work done by the field is entirely converted into heat; we call
this the Smoluchowski point of view, though it is implicit in
\cite{Einstein}. The current
\begin{equation}
j_e=j_\gamma+Vj_c
\end{equation}
of the energy density  $\rho(\Theta+V)$ is conserved, as it obeys
\begin{equation}
\frac{\partial}{\partial t}\rho(\Theta+V)+\mbox{div}j_e=0.
\label{eq-energycons}
\end{equation}
Thus the system obeys the first law of thermodynamics.

The particle current $j_c$ carries with it, by convection,
a heat flow of $\Theta j_c$. This already suggests that the Dufour
effect is to be expected. The surprise here is that the Dufour effect is
$2\Theta j_c$, double what is expected from this intuitive argument. We
call the difference the `anomalous convection'. The Soret effect comes
from the term $-\lambda\rho(1-\rho/\rho_m)\nabla\Theta$ in $j_c$. This
remains non-zero as $\rho_m\rightarrow\infty$, so the effect does not
depend on the presence of a hard core. We see that in this limit, the
Soret coefficient $j_c/(\rho\nabla\Theta)$ is $\lambda$. The fact that in
eq.~(\ref{eq-energycons}) the temperature is added to the external
potential shows that the temperature gradient will cause a flow of
particles; if they are charged, this will be interpreted as the
thermo-electric effect.

The rest of the paper is organised as follows. In \S 2 we outline the model
in discrete space and time. It is a version of the Boltzmann equation,
with discrete energy rather than discrete velocity. It is thus closer to
\cite{Kieg} than \cite{Monaco}. The collision operator is a bistochastic
matrix $T$ conserving energy and particle number (but no other quantities);
it causes transitions (hopping) between particles and holes which are
nearest neighbours. The discrete system is thermodynamically
consistent in its own right, and thus forms the natural discretisation
of the continuum equations of motion eq.~(\ref{eq-continuum}). The state of
the system at any time is described by giving the means of the `slow
variables', here taken to be the particle number, $n=0$ or $n=1$, and the
kinetic energy, at each site. This information defines a
unique grand canonical state at each site; the assumption that this is
the state of the system after a small time-step
is called {\em LTE}, the hypothesis of local thermodynamic equilibrium.
It means that the state is specified by giving
the density and temperature fields, which define a point on our information
manifold, ${\cal M}$. The dynamics of the state in one time-step is
given by applying the map $T$, followed by projection of the resulting
state back to ${\cal M}$.

In \S 3 the continuum limit of the dynamics is taken, with the help of
MAPLE. The size of the lattice spacing $\ell$ and the time step $dt$
are arranged to satisfy $dt=\ell^2$, known as the diffusion limit. This
ensures that the limit exists; it gives the dynamics above.

In \S 4, we verify that the system can be put in Onsager form, though
it is neither linear nor near equilibrium. Onsager symmetry is then
seen to relate the Soret effect to the {\em anomalous} convection, which is
thus the true dual or `inverse' to the Soret effect.

In \S 5 we summarise the paper, and conclude that our simple model,
without any momentum or angular momentum, and with unrealistic density
of states, is able to exhibit the Soret effect and its dual without
the pain of solving a model with interparticle interactions. There is
no reason to expect that these
qualitative conclusions would be altered by a more elaborate model. For
example, we could add a direct transfer of kinetic energy between
occupied sites; this adds further terms to the energy current, and
increases the diagonal part of the Onsager matrix, without affecting
the cross terms (the Soret and Dufour terms).

\section{The Discrete Model with Hard Core}
We start with a finite lattice $\Lambda\subseteq \ell{\bf Z}^\nu$, where
$\nu$ is the dimension of space. A typical point of $\Lambda$ will be
denoted by $x$. At each site $x$ there can be at most one molecule; this
expresses the physical idea that the molecule has a repulsive core of
diameter $\ell$. We take a very simple model of the molecule; it has
no spin or velocity, and its possible states are labelled by $k\in\{0,1,2,
\ldots\}$, denoting the kinetic energy ${\cal K}=k\epsilon$; here $\epsilon
>0$ represents a quantum of energy. This absence of velocity as a label
for the point in phase space leads to a useful simplification of the
mathematics compared with the discrete velocity Boltzmann equation
\cite{Monaco}; we call it the Smoluchowski point of view. The sample space
at $x$ is thus
\begin{equation}
\Omega_x=\{\emptyset,0,1,2,\ldots\}=\{\emptyset\}\cup{\bf N}.
\end{equation}
Here, $\omega_x=\emptyset\in\Omega_x$ means that the site $x$ is empty, and
$\omega_x=k_x\in{\bf N}$ means that the site $x$ is occupied, and that its
kinetic energy is $k_x\epsilon$. We can argue that the fluctuation of the
velocity of a real gas from its local mean value (the drift) is a fast 
variable, and is thermalised in one time-step. In this model,
momentum is not accounted for; however, kinetic energy is.

The sample space, also known as the phase space,
for the system is taken to be
\begin{equation}
\Omega=\prod_{x\in\Lambda}\Omega_x;
\end{equation}
the motivation for choosing this is given in \cite{RFS2}.
The product structure of $\Omega$ ensures that the particles are
indistinguishable, since a configuration is given by a field $\{\omega_x\}_
{x\in\Lambda}$. Thus the Gibbs paradox is avoided from the start, and there
is no need to introduce symmetrising factors.
In a classical stochastic description of such a system, an {\em observable}
is a random variable, that is, a real function on $\Omega$. The set of
all bounded random variables form an algebra, denoted here by ${\cal A}$.
At time $t\in{\bf N}$ the state of the system is described by a
probability measure on $\Omega$, which we denote by $p$; in the theory of
$C^*-$algebras, $p$ is called a normal state on ${\cal A}$. The set of
normal states is a convex set denoted by $\Sigma$.

A useful observable is the number of particles at $x$:
\begin{equation}
{\cal N}_x(\omega)=\left\{\begin{array}{c}
0\mbox{ if }\omega_x=\emptyset,\\
1\mbox{ if }\omega_x\in{\bf N}.\end{array}\right.
\end{equation}
The total number of particles is then the random variable
\begin{equation}
{\cal N}(\omega)=\sum_{\omega\in\Lambda}{\cal N}_x(\omega).
\end{equation}
We introduce an external potential $V(x)$ for the particles. Contrary to
\cite{Chapman2}, we shall not find it necessary to introduce an interaction
between the particles in order to get the Soret and Dufour effects. The
full treatment of a model with interaction leads to a much more complicated
theory than the present one. The `mean field' treatment of the interaction
can be handled as in \cite{Biler}; in this approximation, the interaction
does contribute to the Soret effect, but is not the whole story.

The total energy is the random variable
\begin{equation}
{\cal E}={\cal K}+{\cal V}
\end{equation}
where
\begin{equation}
{\cal V}(\omega)=\sum_xV(x){\cal N}_x(\omega);\hspace{1in}{\cal K}(\omega)
=\sum_x\epsilon k_x{\cal N}_x(\omega),
\end{equation}
$\epsilon>0$ being an energy unit, and
\begin{equation}
\omega=\{\omega_x\}_{x\in\Lambda},\hspace{.5in}\omega_x=\emptyset\mbox{ or }
k_x.
\end{equation}
We are interested in states with finite mean energy
\begin{equation}
p.{\cal E}:=\sum_\omega p(\omega){\cal E}(\omega)<\infty,
\end{equation}
and finite partition function
\begin{equation}
Z=\sum_{\omega}e^{-\beta{\cal E}(\omega)}<\infty\hspace{.6in}\mbox{for all }
\beta>0;
\label{eq-stab}
\end{equation}
this expresses thermodynamic stability.

The function ${\cal E}$ divides $\Omega$ into disjoint energy shells:
\[ \Omega_E=\{\omega\in\Omega:{\cal E}(\omega)=E\},\hspace{.5in}\Omega=
\bigsqcup_{E}\Omega_E.\]
Each $\Omega_E$ must be a finite set, because of eq.~(\ref{eq-stab}).
Similarly we can write
\[\Omega_E=\bigsqcup_n\Omega_{E,n},\]
where
\[\Omega_{E,n}=\Omega_E\cap\Omega_n,\hspace{.4in}\Omega_n=\{\omega\in\Omega
:{\cal N}(\omega)=n\}.\]

The next step in constructing the model is to give a bistochastic map $T$
on $\Sigma$, representing one time-step of the linear part of the dynamics.
This replaces the `collision term' of the full Boltzmann equation. In place
of the large number of collision invariants of the Boltzmann equation, we
require just two conserved quantities, the energy and the particle number.
Now, $T$ is determined by linearity from its action on the point measures
$\delta_{\omega,\omega^\prime}$, and these can be identified with the points
of $\Omega$ \cite{RFS}. To conserve ${\cal E}$ and ${\cal N}$, $T$ must map each
each $\Omega_{E,n}$ to itself. When we give $T$, we are specifying the
conditional probability of transition from $\omega$ to $\omega^\prime$.
We do not attempt to find the most general bistochastic map, but limit
ourselves to the case where
\begin{enumerate}
\item $T_{\omega,\omega^\prime}=T_{\omega^\prime,\omega}$, the symmetric
case.
\item $T_{\omega,\omega^\prime}=0$ if $\omega$ and $\omega^\prime$ differ at
points $x,\;x^\prime\in\Lambda$ which are not nearest neighbours.
\item A transition occurs only by the movement of a particle; there is no
direct transfer of kinetic energy between adjacent particles.
\end{enumerate}
To conserve energy, any change in potential energy in a transition must be
balanced by the opposite change in kinetic energy. Since the latter is an
integral multiple of $\epsilon$, we must suppose that all differences
$V(x)-V(y)$, with $x,\;y\in\Lambda$, are integral multiples of $\epsilon$.
We shall work out the model in detail when $\nu=1$, and $\Lambda$ is an
`interval' of contiguous points in $\ell{\bf Z}$. Thus, for $x\in\Lambda$
with $x+\ell$ also in $\Lambda$, we can write
\begin{equation}
V(x+\ell)-V(x)=\epsilon w_x, \hspace{.5in}\mbox{ with }w_x\in{\bf Z}.
\end{equation}
Because of (2) and (3), above, the possible transitions involve the
movement of a particle from $x$ to a hole at $x+\ell$ or $x-\ell$, or
{\em vice versa}. We shall choose $x$, and consider the case $w_x>0$;
other cases are treated similarly.

Suppose that there is a particle at $x$ and
a hole at $x+\ell$.
In order for the transition $x\mapsto x+\ell$ to be possible, the
kinetic energy of the particle at $x$ must be at least $\epsilon w_x$;
after the transition its kinetic energy is $\epsilon(k_x-w_x)$,
as the particle loses $\epsilon w_x$ in moving uphill. We call $\epsilon
(k_x-w_x)$ the spare kinetic energy of the transition. If there is a
particle at $x+\ell$ and a hole at $x$, the particle can move down to $x$
whatever its kinetic energy happens to be. This move is the inverse
to the first, and to arrive at a symmetric Markov matrix, we choose
the transition rates for these two processes to be the same. We have argued
\cite{RFS3} that a good model for the transition matrix is to choose the
rate to be proportional to the spare kinetic energy, by analogy with
Einstein's hypothesis of stimulated emission. To this, we add one more
unit (the `spontaneous emission') which in the event makes no difference
to the continuum limit.

We see from our answer,
eq.~(\ref{eq-continuum}), that this choice leads to a linear increase in
the thermal conductivity with temperature, roughly in accord with
experiments \cite{Hirschfelder}, p. 574, for various gases in the range
$100^o-300^o\;K.$ It also leads to a diffusion rate that increases linearly,
\cite{RFS3}, as chosen by Smoluchowski.

Let $x\in\Lambda$; let $\omega$ and $\omega^\prime$ be two sample points
on the same $\Omega_{E,n}$, and let $\omega_y^\prime=\omega_y$ for
$y\neq x$ or $x+\ell$. Suppose that $\omega_x=k_x\geq w_x$, and
$\omega_{x+\ell}=\emptyset$, $\omega_x^\prime=\emptyset$ and $\omega_{x+\ell}
^\prime=k_x-w_x$. We call this the hopping criterion, and define the
transition matrix $T_x$ by
\begin{equation}
T_x=
\begin{array}{ccc}
\left.\right.&\omega&\omega^\prime\\
\omega&1-2\lambda\epsilon(k_x-w_x+1)&2\lambda\epsilon(k_x-w_x+1)\\
\omega^\prime&2\lambda\epsilon(k_x-w_x+1)&1-2\lambda\epsilon(k_x-w_x+1)
\end{array}
\end{equation}
It was noticed in \cite{RFS3,RFS4} that transition rates which grow
with
energy need care when the time is discrete, as the transition probability
becomes
larger than 1 for large energies; the stay-as-you-were probability can
become negative. Not only is this unphysical; it leads to unstable
difference equations and spurious chaos \cite{Rondoni}.
It is therefore desirable in our
model to cut off the hopping probability to zero above
\begin{equation}
k_{\rm max}=(2\lambda\epsilon)^{-1}-1.
\label{eq-cutoff}
\end{equation}
By making $\epsilon$ smaller we can make this cut-off as large as we please.
For simplicity of notation, we put the matrix elements of $T_x$ equal to
zero unless $\omega$ and $\omega^\prime$ obey the hopping criterion. That
is, we
remove the diagonal line of units. A matrix like $T_x$ can be defined for
each $x\in\Lambda$. In particular, $T_{x-\ell}$ describes the move down
from $x$ to $x-\ell$, and the move up from $x-\ell$ to $x$, which is
assigned the probability 
\begin{equation}
2\lambda\epsilon(k_{x-\ell}-w_{x-\ell}+1)=2\lambda\epsilon(k_x+1),
\label{fast}
\end{equation}
or zero if $k_{x-\ell}<w_{x-\ell}$.

Both $T_x$ and $T_{x-\ell}$ affect the population at $x$, but other
transitions, $T_y,\;y\neq x,x-\ell$, do not. We define the linear part
of the local dynamics at $x$ to be $(T_x+T_{x-\ell})/2$, and the full linear
dynamics to be given by the symmetric Markov matrix
\begin{equation}
T=\sum_x(T_x+T_{x-\ell})/2.
\end{equation}

The next step in statistical dynamics, the analogue of the Stosszahlansatz,
is the specification of the thermalising map $Q$. This
projects the result of one time-step, $p\mapsto Tp$, onto a subset of
states called the information manifold. We seek to implement mathematically
the physical idea that the kinetic energy at a point $x$ completely
thermalises in one time step. The description of the local state by the
full distribution of the random variable ${\cal K}_x$ is replaced by one
variable, the temperature $\Theta(x)$. According to \cite{Ingarden,Balian}
we have to choose a subspace ${\cal X}$ of slow variables, spanned by
$\{X_0=1,X_1=H,X_2,\ldots,X_N\}$, in the notation of \cite{RFS}. Given a
state $p$, we record the mean values
\begin{equation}
\eta_j=p.X_j,\hspace{.3in}j=1,2,\ldots,N;
\end{equation}
we then define $Qp$ to be the state of maximum entropy having these mean
values. The set of states ${\cal M}({\cal X})=\{Qp:p\in\Sigma\}$ has a
Riemannian structure, and is called the {\em information manifold} of
${\cal X}$. It is parametrised by the means, which are called the
mixture coordinates. If we start with $p\in{\cal M}$, we define one
time-step of the
full (non-linear) dynamics to be $p\mapsto QTp$. This maps ${\cal M}$ to
itself, and is given by difference equations in $N$ variables. This is far
fewer than needed for the linear dynamics.

In the present model, we take the slow variables to be the span of $\{{\cal
N}_x:x\in\Lambda\}\cup\{{\cal K}_x:x\in\Lambda\}$, a total of $N=2|\Lambda|$
variables. In particular the total energy is a slow variable. Let us denote
the means of ${\cal N}_x$ and ${\cal K}_x$ by the fields $n_x$ and $K_x$:
\begin{equation}
n_x=p.{\cal N}_x;\hspace{1in}K_x=p.{\cal K}_x.
\end{equation}
The algebra of observables, ${\cal A}$, is the tensor product
\begin{equation}
{\cal A}=\otimes_x{\cal A}_x
\end{equation}
of local algebras. The state $Qp$ is then the state of maximum entropy with
these means; it is well
known that this is the state in {\em LTE}, the product of local grand
canonical states, independent over $\Lambda$; thus
\begin{equation}
Qp(\omega)=\prod_xp_x(\omega_x)
\label{eq-prod}
\end{equation}
where $p_x$ is the marginal probability of $p$ on $\Omega_x$. In its turn,
$p_x$ is the state on $\Omega_x$ of maximum entropy among states with given
means $n_x,\,K_x$. It is therefore the (fermionic) grand canonical state,
and so has the form, for some parameter $\beta_x$:
\begin{eqnarray}
p_x(\emptyset)&=&1-n_x\\
p_x(k)&=&n_xZ_x^{-1}e^{-\epsilon k\beta_x},
\label{eq-prob}
\end{eqnarray}
where $Z_x=(1-\exp(-\epsilon\beta_x))^{-1}$.
We determine $\beta_x$ or the temperature $\Theta_x=1/(k_B\beta_x)$,
from the mean-value $K_x$:
\begin{eqnarray}
K(x)&=&p.{\cal K}_x=\epsilon n_x\sum_{k=0}kZ_x^{-1}e^{-\epsilon k\beta_x}\\
&=&\frac{\epsilon n_x}{e^{\epsilon\beta_x}-1}.
\end{eqnarray}
Here we have taken the sums up to infinity, neglecting the term
$\exp(-\beta k_{\rm max}\epsilon)$ compared to 1.

We note that
\begin{eqnarray}
n(x)&=&\sum_{k\geq 0}p_x(k)\\
e^{-\epsilon\beta}&=&(1+\epsilon n/K)^{-1}\\
Z&=&1+K/(\epsilon n).
\label{eq-detail}
\end{eqnarray}

Our strategy for specifying one time-step in the dynamics is to start with
$p\in{\cal M}$, and so of the form eq.~(\ref{eq-prod}), (\ref{eq-prob}),
(\ref{eq-detail}), and then find $\hat{p}=Tp$, or rather, its marginals.
From the marginals of $\hat{p}$ we find the new values $\hat{n}(x),
\,\hat{K}(x)$ of the mixture coordinates; from these we can find the new
$\exp(\epsilon\hat{\beta})=1+\epsilon\hat{n}/\hat{K}$. It remains to explain
how to find the marginals of $\hat{p}$.

For a product state over $\Lambda$, eq.~(\ref{eq-prod}), the probability
of the set of points like $\omega$, satisfying the hopping criteria at $x$,
(i.e. $x$ occupied with kinetic energy $\epsilon k_x$, and $x+\ell$ empty)
is $p(x,k)(1-n_{x+\ell})$. Similarly, the set of points like $\omega^\prime$,
(i.e. with $x$ empty, and $x+\ell$ occupied with kinetic energy $\epsilon
(k_x-w_x)$) is $p(x+\ell,k_x-w_x)(1-n_x)$. The other sites $y\neq x,\,x+\ell$
are unaffected by $T_x$ and these factors in eq.~(\ref{eq-prod}) can be
summed over $\omega_y$ to give unity. Thus the marginal at $x$ of $T_xp$ is
\begin{eqnarray}
T_xp(x,k)&=&p(x,k)-p(x,k)\left(1-n_{x-\ell}\right)2\lambda\epsilon\left
(k+1-w_x\right)\nonumber\\
&+&p\left(x+\ell,k-w_x\right)\left(1-n_x\right)2\lambda\epsilon\left(k+1
-w_x\right).
\end{eqnarray}
Similarly
\begin{eqnarray}
T_{x-\ell}p(x,k)&=&p(x,k)-p(x,k)\left(1-n_{x-\ell}\right)2\lambda\epsilon
\left(k_{x-\ell}-w_{x-\ell}+1\right)\nonumber\\
&+&p\left(x-\ell, k_{x-\ell}\right)\left(1-n_x\right)2\lambda\left(k_
{x-\ell}-w_{x-\ell}+1\right).
\end{eqnarray}
Here, $k=k_{x-\ell}-w_{x-\ell}$ must hold, to conserve energy in the
transition. So the change in the marginal at $x$ due to $\left(T_x+T_{x
-\ell}\right)/2$, and therefore to $T$, simplifies a bit to
\begin{eqnarray}
\hat{p}(x,k)&=&p(x,k)-p(x,k)\left(1-n_{x-\ell}\right)\lambda\epsilon
\left(k+1-w_x\right)\nonumber\\
&+&p\left(x+\ell,k-w_x\right)\left(1-n_x\right)\lambda\epsilon\left(k+1-w_
x\right)\nonumber\\
&-&p(x,k)\left(1-n_{x-\ell}\right)\lambda\epsilon(k+1)\nonumber\\
&+&p\left(x-\ell,k+w_{x-\ell}\right)\left(1-n_x\right)\lambda\epsilon(k+1).
\end{eqnarray}
From this we can find the new values, the one-step updates, $\hat{n},
\,\hat{K}$:
\begin{eqnarray}
\hat{n}_x&=&\sum_{k\geq 0}\hat{p}(x,k)\nonumber\\
&=&n_x-\lambda\epsilon n_x\left(1-n_{x+\ell}\right)Z_x^{-1}\sum_{k\geq w_x}
\left(k+1-w_x\right)e^{-\epsilon k\beta_x}\nonumber\\
&+&\lambda\epsilon n_{x+\ell}\left(1-n_x\right)Z_{x+\ell}^{-1}\sum_{k\geq
w_x}\left(k+1-w_x\right)e^{-\epsilon\left(k-w_x\right)\beta_{x+\ell}}
\nonumber\\
&-&\lambda\epsilon n_x\left(1-n_{x-\ell}\right)Z_x^{-1}\sum_{k\geq 0}
(k+1)e^{-\epsilon k\beta_x}\nonumber\\
&+&\lambda\epsilon n_{x-\ell}\left(1-n_x\right)Z_{x-\ell}^{-1}\sum_{k\geq0}
(k+1)e^{-\epsilon\left(k+w_{x-\ell}\right)\beta_{x-\ell}},
\label{eq-n}
\end{eqnarray}
and
\begin{eqnarray}
\hat{K}_x&=&\sum_{k\geq 0}\epsilon k\hat{p}(x,k)\nonumber\\
&=&K_x-\epsilon^2\lambda n_x\left(1-n_{x+\ell}\right)Z_x^{-1}\sum_{k\geq
w_x}k\left(k+1-w_x\right)e^{-\epsilon k\beta_x}\nonumber\\
&+&\epsilon^2\lambda n_{x+\ell}\left(1-n_x\right)Z_{x+\ell}^{-1}\sum_{k\geq
w_x}k\left(k+1-w_x\right)e^{-\epsilon\left(k-w_x\right)k\beta_{x+\ell}}\nonumber\\
&-&\epsilon^2\lambda n_x\left(1-n_{x-\ell}\right)Z_x^{-1}\sum_{k\geq0}
k(k+1)e^{-\epsilon k\beta_x}\nonumber\\
&+&\epsilon^2\lambda n_{x-\ell}\left(1-n_x\right)Z_{x-\ell}^{-1}\sum_{k\geq
0}k(k+1)e^{-\epsilon\left(k+w_{x-\ell}\right)\beta_{x-\ell}}.
\label{eq-K}
\end{eqnarray}
The sum should go up to the largest $k$ consistent with the positivity
of the diagonal matrix element of $T_{\omega,\omega^\prime}$ at $x$ and
$x-\ell$. This ensures that the model is entropy increasing,
and that the discrete dynamics is stable.

The dynamics is the explicit map $n\mapsto\hat{n},\,K\mapsto\hat{K}$,
and from $\hat{K}$ we can compute $\hat{\beta}$, and thus get the new
point of ${\cal M}$. In the next section we take the continuum limit of this
dynamics. The numerical solution of the resulting reaction-diffusion
equations might be best done using these discrete equations: energy is
conserved and entropy increases in all approximations, which ensures that
the numerical solutions have reasonable physical properties, and do not
exhibit spurious chaos, (chaos not present in the continuum equations).

\section{The Continuum Limit}
We now take the continuum limit of the updating equations, thus:
\begin{eqnarray}
dt&=&\ell^2\rightarrow 0;\hspace{.4in}\epsilon=\gamma\ell\rightarrow 0
\hspace{.4in}(\gamma\mbox{ fixed})\\
\frac{\hat{K}-K}{dt}&\rightarrow&\frac{\partial K}{\partial t};\hspace{.4in}
\frac{1}{\ell}\rightarrow\rho_m;\hspace{.4in}\rho=\frac{n}{\ell}\\
\frac{V(x+\ell)-V(x)}{\ell}&\rightarrow&\frac{\partial V}{\partial x}
\hspace{.4in}\mbox{and so on}.
\end{eqnarray}
This is achieved by writing
\begin{eqnarray}
\beta_{x\pm \ell}&=&\beta\pm \ell\beta1+(1/2)\ell^2\beta2\\
n_{x\pm\ell}&=&n\pm\ell n1+(1/2)\ell^2n2\\
\epsilon w_{x-\ell}&=&V(x)-V(x-\ell)=\ell V1-(1/2)\ell^2V2\\
\hat{n}-n&=&A1+A2+A3+A4.
\end{eqnarray}
Here, $\beta1$ is first the derivative of $\beta$, $\beta2$ is the
second derivative, and the same for the functions $n$ and $V$; the
A's are four expressions in eq.~(\ref{eq-n}):
\begin{eqnarray}
A1&=&-\epsilon\lambda n(1-n3)e^{-\epsilon w\beta}
\left(1-e^{-\epsilon\beta}\right)\sum_{k\geq0}(k+1)e^{-\epsilon\beta k}\\
A2&=&\epsilon\lambda n3(1-n)\left(1-e^{-\epsilon\beta3}\right)
\sum_{k\geq0}(k+1)e^{-\epsilon\beta3\,k}\\
A3&=&-\epsilon\lambda n(1-n4)\left(1-e^{-\epsilon\beta}
\right)\sum_{k\geq0}(k+1)e^{-\epsilon\beta k}\\
A4&=&\epsilon\lambda n4(1-n)\left(1-e^{-\epsilon\beta4}\right)
e^{-\epsilon\beta4w4}\sum_{k\geq0}(k+1)e^{-\epsilon\beta4\,k},
\end{eqnarray}
where $\beta3=\beta_{x+\ell}$, $\beta4=\beta_{x-\ell}$, $n3=n_{x+\ell}$,
$n4=n_{x-\ell}$, $w4=w_{x-\ell}$.
The sum is allowed to go to infinity in the continuum limit.

We ask MAPLE to evaluate $\hat{n}-n$ to lowest non-vanishing order in
$\ell$, namely, $\ell^2$; we substitute $\epsilon=\gamma\ell$, with
$\gamma$  fixed, $n3=n+\ell n^\prime+(1/2)\ell^2n^{\prime\prime}$ etc.,
$\beta(x)=1/\Theta(x)$, and $\hat{n}-n$ by $\ell^2\partial n/\partial t$.
In one dimension the result, as $\ell\rightarrow0$, is
\begin{equation}
\frac{\partial n}{\partial t}+\mbox{div}\,j_n=0
\end{equation}
where
\begin{equation}
j_n=-\lambda\left(\Theta n^\prime+\left(\Theta^\prime+V^\prime\right)n(1-n)
\right).
\label{eq-eq}
\end{equation}
We see that this is independent of $\gamma$.

The particle density is $n/\ell=\rho$; then $n=\rho\ell$ which we replace 
by $\rho/\rho_m$ rather than by zero in the limit. So by dividing
eq.~(\ref{eq-eq}) by $\ell$ gives our equation of motion for the particle
density:
\begin{equation}
j_c=-\lambda\left(\Theta\nabla\rho+\rho(1-\rho/\rho_m)\nabla
(\Theta+V)\right).
\end{equation}
The term $\lambda\rho(1-\rho/\rho_m)\nabla\Theta$ is the thermal
diffusion; if $\rho\ll\rho_m$ we see that the Soret coefficient is
exactly $\lambda$, but that for dense fluids (near solidification)
it is smaller. The Soret term is absent in the model of dense fluids
presented in \cite{RFS2}. This arises because there the particles do not
carry heat, unlike those of the present model.

For the kinetic energy density $K(x)$, we write
\begin{equation}
\hat{K}-K=A1+A2+A3+A4,
\end{equation}
where
\begin{eqnarray}
A1&=&-\epsilon^2\lambda n(1-n3)\left(1-e^{-\epsilon\beta}\right)e^{-\epsilon
\beta w}\sum_{k\geq0}(k+w)(k+1)e^{-\epsilon\beta k}\\
A2&=&\epsilon^2\lambda(1-n)n3\left(1-e^{-\epsilon\beta3}\right)\sum_{k\geq0}
(k+w)(k+1)e^{-\epsilon\beta3\,k}\\
A3&=&-\epsilon^2\lambda n(1-n4)\left(1-e^{-\epsilon\beta}\right)\sum_{k\geq0}
k(k+1)e^{-\epsilon\beta k}\\
A4&=&\epsilon^2\lambda n4(1-n)\left(1-e^{-\epsilon\beta4}\right)e^{-\epsilon
\beta4\,w4}\sum_{k\geq0}k(k+1)e^{-\epsilon\beta4\,k}.
\end{eqnarray}
Now we put $K=n\Theta$ and take the limit $\ell\rightarrow0$; MAPLE gives
\begin{eqnarray}
\frac{\partial(n\Theta)}{\partial t}&=&-\lambda
\left\{\rule{0mm}{5mm}n^2(V^\prime)^2-
3V^\prime n^\prime\Theta-3V^\prime n\Theta^\prime-n(V^\prime)^2\right.
\nonumber\\
&+&4n^2(\Theta^\prime)^2
+4n^2\Theta\Theta^{\prime\prime}-4n\Theta\Theta^{\prime\prime}+2n^2\Theta
V^{\prime\prime}\nonumber\\
&+&3n^2V^\prime\Theta^\prime+8n\Theta\Theta^\prime
n^\prime+4nn^\prime V^\prime\Theta\nonumber\\
&-&\left.\rule{0mm}{5mm}2n^{\prime\prime}\Theta^2-8n^\prime\Theta\Theta^\prime-2n\Theta V^{
\prime\prime}-4n\left(\Theta^\prime\right)^2\right\}\nonumber\\
&=&-j_nV^\prime-2\mbox{div}\,\left(\Theta j_n\right)+2\lambda\mbox{div}\,
\left(\Theta(1-n)n\Theta^\prime\right).
\label{eq-rhothetadot}
\end{eqnarray}
Again we see that $\gamma$ drops out. Putting $n/\ell=\rho$ and
$1/\ell=\rho_m$ we get in $\nu$-dimensions:
\begin{equation}
\frac{\partial(\rho\Theta)}{\partial t}=-j_c.\nabla V-2\mbox{div}\,
\left(\Theta j_c\right)+2\lambda\mbox{div}\left(\Theta\rho(1-\rho/\rho_m)
\nabla\Theta\right).
\end{equation}
Thus, at least at a formal level, the equations discussed in \S 1 
are `derived' from a discrete stochastic model.
Our derivation has not proved that the solutions to the discrete equations
converge to solutions of the nonlinear coupled equations in the continuum.
Indeed these are not uniformly elliptic, and a proof of existence of
even local solutions needs care. Some results for similar systems are
presented in \cite{RFS5,Biler,Biler2}.

\section{The Onsager Form}
We now argue that only the `anomalous' part of the Dufour effect is the
true Onsager dual to the Soret effect. We see this in the model of
\cite{RFS2}; this model has no Soret effect, but the heat current does
contain the normal convection term $\Theta j_c$. In the classical Boltzmann
equation the heat content of a fluid is $3/2\Theta\rho$, instead of our
$\Theta\rho$; we would expect, then, a convection term $3/2\Theta j_c$
in the heat current. However, in \cite{Chapman2} the term $5/2\Theta j_c$
is found. We see that of this, $3/2\Theta j_c$ is normal convection,
and $\Theta j_c$ is anomalous convection; the latter is the same as in
the present model. We shall show that Onsager symmetry relates the Soret
effect to the anomalous part of the convection.

Both the model of \cite{RFS2}, and the present model, have the same sample
space, slow variables and conserved quantities, namely the particle number
and energy. It follows from the argument given in
\cite{RFS3} that both models have the same entropy and thermodynamic
forces. They differ in their dynamics, which in Onsager theory
is determined by the way the currents depend on the forces. We first
find the entropy of the discrete model, and then take its formal
continuum limit. The entropy of the state
\begin{equation}
p(\omega)=\prod_x\{(1-{\cal N}_x(\omega))(1-n_x)+{\cal N}_x(\omega)p_x(\omega_x)\}
\end{equation}
which is independent over $\Lambda$, is the sum of the contributions at
each $x$; thus
\begin{eqnarray}
S(p)&=&-\sum_\omega p(\omega)\log p(\omega)\nonumber\\
&=&-\sum_x\left(1-n_x\right)\log\left(1-n_x\right)-\sum_xn_x\sum_{k\geq0}
p_x(k)\log p_x(k)\nonumber\\
&=&-\sum_x\left(1-n_x\right)\log\left(1-n_x\right)-\sum_xn_x\log n_x
\nonumber\\
&+&\sum_x n_x\left\{\log Z_{\beta_x}+\beta_x\epsilon/\left(e^{\epsilon
\beta_x}-1\right)\right\}
\end{eqnarray}
Now,
\begin{eqnarray}
-\sum_xn_x\log n_x&-&\sum_x\left(1-n_x\right)\log\left(1-n_x\right)\\=
-\sum_x\frac{n_x}{\ell}\left(\log\frac{n_x}{\ell}+\log \ell\right)
\ell&-&
\sum_x\left(\frac{1}{\ell}-\frac{n_x}{\ell}\right)\log\left(1-n_x\right)
\end{eqnarray}
and in the continuum limit, $\ell\rightarrow0$, $\sum_x\ell$ becomes
$\int\,dx$, $n_x/\ell$ becomes $\rho(x)$ and $1/\ell$ becomes $\rho_m$.
So the dichotomic part of the entropy becomes the differential entropy
\begin{equation}
-\int\rho(x)\log\rho(x)\,dx-\int\left(\rho_m-\rho\right)\log\left(1-
\rho/\rho_m\right)dx,
\end{equation}
apart from the large positive term
\begin{equation}
-\log\ell\int\rho(x)\,dx.
\end{equation}
However, this term can be dropped; for $\int\rho(x)\,dx$ is constant
in time, so the divergent term does not contribute to $dS/dt$.
The Gibbsian term in $S_x$ also has a simple limit:
\begin{eqnarray}
n_x\left(\log Z_{\beta_x}+\beta_xK_x\right)&=&n_x\left(-\log\left(
1-e^{-\epsilon\beta_x}\right)+\beta_x\epsilon/\left(e^{\epsilon\beta}-1\right)
\right)\nonumber\\
&=&-n_x\log\epsilon-n_x\log\beta_x+n_x+O(\epsilon).
\end{eqnarray}
Again, summing over $x$ leads to the divergent but constant terms
$(-\log\epsilon+1)\sum_xn_x$, which can be dropped, leaving
\[-\sum_x\log\beta_x n_x\rightarrow\int\rho(x)\log\Theta(x)\,dx;\]
The continuum entropy is therefore
\begin{eqnarray*}
S&=&-\int\rho(x)\log\rho(x)\,dx-\int\left(\rho_m-\rho\right)\log
\left(1-\rho/\rho_m\right)\,dx\\
&+&\int\rho\log\Theta(x)\,dx,
\end{eqnarray*}
as claimed in \cite{RFS2}, eq.~(49).

The next step \cite{RFS3} in setting up a comparison with Onsager theory is
to write $\dot{S}$ as a (continuous version) of Onsager's ansatz
\cite{Groot}
\begin{equation}
\Theta\dot{S}=\sum_\alpha X^\alpha j_\alpha.
\end{equation}
We use the identities
\begin{eqnarray}
\int\rho\dot{\Theta}/\Theta\,dx&=&\int\dot{\left(\rho\Theta\right)}/
\Theta\,dx\\
\int\rho\dot{\rho}/\rho\,dx&=&\int\left(\rho_m-\rho\right)\dot{\rho}/
\left(\rho_m(\rho_m-\rho)\right)\,dx=0
\end{eqnarray}
to get
\begin{eqnarray}
\dot{S}&=&-\int\dot{\rho}\log\rho\,dx+\int\dot{\rho}\log\left(1-\rho/
\rho_m\right)\,dx\nonumber\\
&+&\int\dot{\rho}\log\Theta\,dx+\int\dot{(\rho\Theta)}/
\Theta\,dx.
\label{eq-sdot}
\end{eqnarray}
In both the present model and that of \cite{RFS2} there are two
conserved densities, the particle number $\rho$ and the energy ${\cal E}=
\rho(V+\Theta)$; thus $\dot{\rho\Theta}=\dot{\cal E}-\dot{\rho}V$.
We impose the condition of `no flow' on the boundary,
$\partial\Lambda$, which could be at infinity. Thus the components of
the current normal to the boundary, $j_c^\perp$ and $j_e^\perp$, vanish
on $\partial\Lambda$. In both models, then, we use the conservation laws
and then integrate by parts in eq.~(\ref{eq-sdot})
and discard the boundary term, to get:
\begin{eqnarray}
\dot{S}&=&-\int\dot{\rho}\left(\log(\rho\rho_m/(\rho_m-\rho))-\log\Theta
\right)+\int\left(\dot{\cal E}-\dot{\rho}V\right)/\Theta\,dx\\
&=&-\int j_c.\nabla\left(\log(\rho/(1-\rho/\rho_m))-\log\Theta+V/\Theta
\right)\,dx\nonumber\\
&+&\int j_e.\nabla(1/\Theta)\,dx.
\end{eqnarray}
This is true whatever the equation of motion, provided that the particle
number and energy are conserved. In particular, we can vary the hopping
probability, $\lambda$, as a function of $(x,t)$, so that $j_c$ and $j_e$
are arbitrary functions. So from the Onsager ansatz for $\dot{S}$ we can
read off the thermodynamic forces
\begin{eqnarray}
\frac{X^c}{\Theta}&=&-\nabla\left\{\log\left(\frac{\rho}{1-\rho/\rho_m}
\right) +\frac{V}{\Theta}-\log\Theta\right\}\\
&=&-\frac{1}{\rho\left(1-\rho/\rho_m\right)}\left\{\nabla\rho+
\frac{\nabla V}{\Theta}\rho\left(1-\rho/\rho_m\right)\right\}-(V+\Theta)
\nabla\left(\frac{1}{\Theta}\right);\hspace{.2in}\\
\frac{X^e}{\Theta}&=&\nabla\frac{1}{\Theta}.
\label{eq-forces}
\end{eqnarray}
These expressions for the thermodynamic forces conjugate to the currents
$j_c$ and $j_e$ hold for both models; we consider them in turn.
In the model of \cite{RFS2} the equations of motion are
\begin{eqnarray}
j_c&=&-\lambda\left\{\nabla\rho+\rho\left(1-\rho/\rho_m\right)\frac{\nabla
V}{\Theta}\right\}\\
j_e&=&-\lambda\left\{\rho\left(1-\rho/\rho_m\right)\nabla\Theta\right\}
+(\Theta+V)j_c.
\end{eqnarray}
This model has no Soret effect, but shows the `normal' convection, in our
terminology, because of the contribution $\Theta j_c$ to the heat current.
This will be observed as the Dufour effect, since there is a heat flow
if $\nabla\Theta=0$ but $\nabla\rho\neq0$. Somewhat luckily (for Onsager
theory), the currents are linear expressions in $X^c$ and $X^e$,
with no derivatives, but with nonlinear coefficients:
\begin{eqnarray}
j_c&=&\lambda\rho\left(1-\rho/\rho_m\right)\frac{X^c}{\Theta}+\lambda
(V+\Theta)\rho\left(1-\rho/\rho_m\right)\frac{X^e}{\Theta};\hspace{.2in}\\
j_e&=&(V+\Theta)\lambda\rho\left(1-\rho/\rho_m\right)\frac{X^c}{\Theta}\nonumber\\
&+&\lambda\left\{\lambda\rho\left(1-\rho/\rho_m\right)\Theta^2+(\Theta+V)^2
\rho\left(1-\rho/\rho_m\right)\right\}\frac{X^e}{\Theta}.\hspace{.2in}
\end{eqnarray}
We see that Onsager symmetry holds, and that the Onsager matrix is positive
definite. Hence entropy is an increasing function of time, as expected.

In the present model, we see that the currents, given in
eq.~(\ref{eq-continuum}), can be expressed in terms of the {\em same}
thermodynamic forces eq.~(\ref{eq-forces}) thus:
\begin{eqnarray}
j_c&=&\lambda\rho\left(1-\rho/\rho_m\right)\frac{X^c}{\Theta}+\lambda(V+2
\Theta)\rho\left(1-\rho/\rho_m\right)\frac{X^e}{\Theta}\\
j_e&=&(V+2\Theta)\lambda\rho\left(1-\rho/\rho_m\right)\frac{X^c}{\Theta}
\nonumber\\
&+&\lambda\left\{(V+2\Theta)^2\rho\left(1-\rho/\rho_m\right)+2\Theta^3\rho
\left(1-\rho/\rho_m\right)\frac{X^e}{\Theta}\right\}.
\end{eqnarray}
Again, Onsager symmetry and positivity hold. The anomalous part of the
convection, the factor `2' in $V+2\Theta$, can be inferred using Onsager
symmetry and the Soret term, $\nabla\Theta$, in $j_c$. In our approach, all these properties, the Soret effect, the
anomalous Dufour effect and the Onsager symmetry, follow from the model,
rather than being put in, as in the Onsager theory.

We can relate the thermodynamic forces to the gradients of the canonical
coordinates $\xi^\alpha(x)$ of the information manifold (when the states
are independent over $\Lambda$), according to the general theory; see
eq.~(15) of \cite{RFS3}. We shall now verify that this is true in this
model, in the continuum limit. Recall that in the discrete model we
write the density matrix as
\begin{equation}
\rho=e^{-\sum_0\xi^\alpha H_\alpha}.
\end{equation}
Here, $H_0=1$, $\xi^0=\log\Xi$, where $\Xi$ is the grand partition function;
the $H_\alpha$ are the conserved densities. In our model, these
are the number of particles at $x$, ${\cal N}_x$, and the energy at $x$,
${\cal E}_x={\cal V}_x+{\cal K}_x$. A point on the
information manifold, for finite $\Lambda$, is the product over $x$
of a probability which we write in the usual form of a local equilibrium
state:
\begin{equation}
p(x)=e^{-\log\Xi}e^{-\beta_x({\cal E}_x-\mu_x{\cal N}_x)}.
\end{equation}
We can thus identify the canonical coordinates as $\xi_x^c=-\beta_x\mu_x$
and, as expected, $\xi_x^e=\beta_x$. To relate $\xi^c$ to $n$, 
one of the mixture coordinates, note that (omitting the label $x$)
\begin{equation}
\Xi^{-1}=1-n
\label{eq-forn}
\end{equation}
and
\begin{eqnarray}
\Xi&=&\sum_{\omega}e^{\beta\mu{\cal N}}e^{-\beta{\cal E}}\\
&=&1+e^{\beta(\mu-V)}\sum_{k\geq0}e^{-\epsilon\beta k}\nonumber\\
&=&1+\frac{e^{\beta(\mu-V)}}{1-e^{-\epsilon\beta}}.
\label{eq-forxi}
\end{eqnarray}
We combine eq.~(\ref{eq-forn}) with eq.~(\ref{eq-forxi}) to get
\begin{eqnarray}
\beta\mu&=&\beta V+\log\left(n/(1-n)\right)+\log\left(1-e^{-\epsilon\beta}
\right)\\
&=&\beta V+\log\frac{\rho}{\rho-\rho/\rho_m}+\log\epsilon-\log\beta+
O(\epsilon)
\end{eqnarray}
for small $\epsilon$. In the continuum limit we drop the infinite constant
$\log\epsilon$, as only the gradient is used. Thus we get
\begin{equation}
\xi^c=-\beta\mu=-\log\frac{\rho}{1-\rho/\rho_m}+\log\Theta-\frac{V}{\Theta}
\end{equation}
and $X^c/\Theta=\nabla\xi^c$, using eq.~(70).
This, together with
eq.~(72) shows that $(\xi^e,\xi^c)$ are ``potentials'' for
$(-X^e/\Theta,-X^c/\Theta)$.

\section{Conclusion}
We have constructed an example of nonequilibrium thermodynamics
obeying the first and second laws, and which exhibits the Soret and
Dufour effects. Apart from the hard core, no interparticle potential is
postulated, and indeed the effects persist in the limit $\rho_m\rightarrow
\infty$, corresponding to no hard core. This should be compared with
the kinetic theory described in \cite{Chapman2}, p.103; this gives the
impression that the Soret and Dufour effects depend on the careful
inclusion of interparticle forces and that they are present only in
gas mixtures. In our approach the effect of the interparticle forces are
included only indirectly, in that the dynamics includes the thermalising
map $Q$, which ensures that the motion is confined to the information
manifold. Thus, whatever the forces are, they keep the system in local
thermodynamic equilibrium. This leads to a simple understanding of the
effects; regions of higher temperature contain more high-speed molecules
than regions of low temperature, and so they preferentially move from high
to low temperatures. In our model there is no velocity variable, and by
high speed we mean molecules that hop with greater probability. In
\cite{Chapman2} an abnormal convection is also found, and is interpreted
as ``convection of enthalpy''. It is the
differential Soret effect between isotopes
that is emphasised there. This is the reason for the great industrial
importance of the effect; it allows gas mixtures to be separated by
a heat gradient. To get this effect we would need to allow the hopping
rate $\lambda$ to depend on the mass of the molecule.

We have remarked that the density of states in our model is unity; in
a semiclassical model in one dimension, there are two states of each energy,
corresponding to the two directions of motion, so apart from this
trivial factor, it would appear that our model is one dimensional.
However, the hopping rate is not that of a particle in one dimension.
In the free dynamics
between collisions the rate of movement is proportional to the velocity i.e.
$E^{1/2}$, whereas
in our model it is proportional to the spare kinetic energy $E$. 
This larger hopping rate partially compensates for the lack of multiplicity
of states, as can be seen by the following argument.
In three dimensions the number of states of a single particle in the
semiclassical model
is proportional to the volume of phase space, and thus has a factor
$p^2\,dp$ (where $p$ is now the momentum) which is proportional to $E^{1/2}dE$.
Suppose a particle in three dimensions at a lattice site $x\in\Lambda$
having momentum $p$ hops to a neighbouring site, $x+\ell$ against the force
given by the gradient of the
potential $V$. Its momentum after the transition will be slightly different,
say $p^\prime$, which is determined by $p$ and the force. Thus each of the
many states at $x$ can make a momentum-conserving transition to only one
of the many states of the right energy at $x+\ell$. So the number of states
making the transition is proportional to the number of states, namely
$E^{1/2}$. The rate of flow of such an element of phase space is
proportional to $E^{1/2}$ as well, so the number making the transition is
$En(E)$, where $n(E)$ is the occupation number. This is the same as that
due to stimulated hopping, as assumed in the present paper in
eq.~(\ref{fast}). As we saw, the
extra hop due to spontaneous hopping did not contribute to the continuum limit. 
It should be mentioned that this does
not mean that model of the present paper gives the same answer (apart from
trivial factors)
as a three-dimensional model with the``correct'' multiplicity, and rates 
proportional to the speed. Indeed, the factors $Z^{-1}$ in the rate
equations eq.~(\ref{eq-n}) and (\ref{eq-K}) also depend on the multiplicity.
These factors are proportional to
$\Theta^{-3/2}$ instead of $\Theta^{-1}$ as in the present paper.

We can also regard the present model as describing a dense liquid
of complex atoms, which have little kinetic energy, but which have
a large number of excited states modelled by the levels $\epsilon k,\;\;
k=0,1,2,\ldots,k_{\rm max}$. As remarked, to get the equations (1) and
(2), we approximate $(1-\exp(-\beta\epsilon k_{\rm max}))$ by 1,
which requires that the temperature is low enough so that the
states of high energy are not excited much.
The density of states
can be independently checked by experiment. For such a liquid,
it would be interesting to see whether it is true that
Dufour effect is double the convection (the factor $2$ in eq.~(2)).

Our model is more in the spirit of
the discrete energy
Boltzmann equations \cite{Kieg} than the discrete velocity models
\cite{Monaco} which conserve momentum as well as energy. For example,
\cite{Droz} considers
a lattice gas version of the Boltzmann equation, with a collision term
between pairs of particles. The Dufour effect is predicted.
In \cite{Droz}, only two speeds occur. The number of particles having each
speed is separately conserved, so there is no thermal mixing in the
scattering. The up-date equations for the densities are of the eighth
degree. The authors compare numerical simulations of the exact
model with solutions
to a simplified model in which the momentum is put equal to zero,
and a thermalising assumption is imposed. They remark that the simplified
model is very close to the simulations. Our result shows that the Soret
effect as well as the Dufour effect can be obtained without the two-body
scattering.

We may rather easily vary the
lattice shape, and allow hopping to next-nearest neighbours; we tried
several such variants, and got the same continuum limit provided that
we adjusted $\lambda$ so that the finite-difference operator in the
transition matrix $T$ approximates the Laplacian. So the limit is rather
robust.

\noindent{\bf Acknowledgements}\\
This work was started at the university of Madeira, Summer, 1997; I
thank Prof. Benilov
for the hospitality of the Dept. of Physics, and H. Nencka for arranging
the visit. It was completed Feb 1998, in the Dept. of Mathematics,
politecnico of Turin; I thank G. Pistone for the invitation and L. Rondoni
for arranging the visit.

\section{The Maple V Program}
\subsection{The Density}
$\sharp$We use the notation $n1=dn/dx,\;\;n2=d^2n/dx^2$,\\
$\sharp$  n3=n(x+l), n4=n(x-l), and the same for\\
$\sharp$ beta and w.\\
assume(epsilon$>$0);assume(beta$>$0);\\
A1:=-lambda*epsilon*n*(1-n3)*exp(-epsilon*beta*w)*(1-exp(-epsilon*beta))*\\
sum((k+1)*exp(-epsilon*beta*k),k=0..infinity);\\
B1:=simplify(A1);\\
C1:=series(B1,epsilon,5);\\
D1:=convert(C1,polynom);\\
F1:=simplify(D1);\\
G1:=subs(epsilon=l*gamma,F1);\\
H1:=subs(n3=n+l*n1+(1/2)*l\begin{math}\widehat{\rule{3mm}{0mm}}\end{math}
2*n2,G1);\\
J1:=series(H1,l,3);\\
Z1:=convert(J1,polynom);\\
assume(beta3$>$0);\\
A2:=epsilon*lambda*n3*(1-n)*(1-exp(-beta3*epsilon))*sum((k+1)*\\
exp(-beta3*epsilon*k),k=0..infinity);\\
B2:=series(A2,epsilon,5);\\
C2:=convert(B2,polynom);\\
D2:=simplify(C2);\\
F2:=subs(epsilon=l*gamma,D2);\\
G2:=subs(n3=n+l*n1+(1/2)*l\begin{math}\widehat{\rule{3mm}{0mm}}\end{math}2
*n2,F2);\\
H2:=subs(beta3=beta+l*beta1+(1/2)*l\begin{math}\widehat{\rule{3mm}{0mm}}
\end{math}2*beta2,G2);\\
J2:=series(H2,l,3);\\
K2:=convert(J2,polynom);\\
Z2:=simplify(K2);\\
A3:=-epsilon*lambda*n*(1-n4)*(1-exp(-beta*epsilon))*sum((k+1)*\\
exp(-beta*epsilon*k),k=0..infinity);\\
B3:=series(A3,epsilon,5);\\
C3:=convert(B3,polynom);\\
D3:=subs(epsilon=l*gamma,C3);\\
F3:=subs(n4=n-l*n1+(1/2)*l\begin{math}\widehat{\rule{3mm}{0mm}}\end{math}
*n2,D3);\\
G3:=series(F3,l,3);\\
H3:=convert(G3,polynom);\\
Z3:=simplify(H3);\\
assume(beta4$>$0);\\
A4:=epsilon*lambda*n4*(1-n)*(1-exp(-beta4*epsilon))*\\
exp(-beta4*epsilon*w4)*sum((k+1)*exp(-beta4*epsilon*k),k=0..infinity);\\
B4:=series(A4,epsilon,5);\\
C4:=convert(B4,polynom);\\
D4:=subs(epsilon=gamma*l,C4);\\
F4:=subs(n4=beta-l*n1+(1/2)l\begin{math}\widehat{\rule{3mm}{0mm}}\end{math}
2*n2,D4);\\
G4:=subs(beta4=beta-l*beta1+(1/2)*l\begin{math}\widehat{\rule{3mm}{0mm}}
\end{math}2*beta2,F4);\\
H4:=subs(w4=w-l*w1+(1/2)*l\begin{math}\widehat{\rule{3mm}{0mm}}
\end{math}2*w2,G4);\\
J4:=series(H4,l,3);\\
K4:=convert(J4,polynom);\\
Z4:=simplify(K4);\\
ans1:=simplify(Z1+Z2+Z3+Z4);\\
ans2:=subs(beta=1/Theta(x),ans1);\\
ans3:=subs(beta1=diff(1/Theta(x),x),ans2);\\
ans4:=subs(beta2=diff(1/Theta(x),x\$2),ans3);\\
ans5:=subs(w=diff(V(x),x),ans4);\\
ans6:=subs(w1=diff(V(x),x\$2),ans5);\\
ans:=simplify(ans6);

Simple manipulation then gives eq.~(\ref{eq-ncurrent}).

\subsection{The Heat}
$\sharp$ Soret2.ms; as before, we take the four terms from A to Z.\\
$\sharp$ beta1 denotes $d\beta/dx$, beta2 denotes $d^2\beta/dx^2$,\\
$\sharp$ beta3 denotes $\beta(x+\ell)$; beta4 denotes $\beta(x-\ell)$.\\
$\sharp$ and the same for n and w.\\
assume(beta$>$0); assume(epsilon$>$0);\\
A1:=-lambda*epsilon$\widehat{\rule{3mm}{0mm}}$2*n*(1-n3)*exp(-beta*epsilon
*w)*\\
(1-exp(-beta*epsilon))*sum((k+w)*(k+1)*exp(-beta*epsilon*k),k=0..infinity);\\
B1:=simplify(A1);\\
C1:=series(B1,epsilon,6);\\
D1:=convert(C1,polynom);\\
F1:=subs(epsilon=l*gamma,D1);\\
G1:=subs(n3=n+l*n1+(1/2)*l$\widehat{\rule{3mm}{0mm}}$2*n2,F1);\\
H1:=series(G1,l,3);\\
J1:=convert(H1,polynom);\\
Z1:=simplify(J1);\\
assume(beta3$>$0);\\
A2:=epsilon$\widehat{\rule{3mm}{0mm}}$2*beta*(1-n)*n3*(1-exp(-epsilon*beta3))*
sum((k+w)*(k+1)*exp(-epsilon*beta3*k),k=0..infinity);\\
B2:=simplify(A2);\\
C2:=series(B2,epsilon,6);\\
D2:=convert(C2,polynom);\\
F2:=subs(epsilon=l*gamma,D2);\\
G2:=subs(n3=n+l*n1+(1/2)*l$\widehat{\rule{3mm}{0mm}}$2*n2,F2);\\
H2:=subs(beta3=beta+l*beta1+(1/2)*l$\widehat{\rule{3mm}{0mm}}$2*beta2,G2);\\
J2:=series(H2,l,3);\\
K2:=convert(J2,polynom);\\
Z2:=simplify(K2);\\
A3:=-epsilon$\widehat{\rule{3mm}{0mm}}$2*lambda*n*(1-n4)*(1-exp(-epsilon*beta))*sum(k*(k+1)*\\
exp(-epsilon*beta*k),k=0..infinity);\\
B3:=simplify(A3);\\
C3:=series(B3,epsilon,6);\\
D3:=convert(C3,polynom);\\
F3:=subs(epsilon=l*gamma,D3);\\
G3:=subs(n4=n-l*n1+(1/2)*l$\widehat{\rule{3mm}{0mm}}$2*n2);\\
H3:=series(G3,l,3);\\
K3:=convert(H3,polynom);\\
Z3:=simplify(K3);\\
assume(beta4$>$0);\\
A4:=epsilon$\widehat{\rule{3mm}{0mm}}$2*lambda*n4*(1-n)*(1-exp(-epsilon*beta4))*\\
exp(-epsilon*beta4*w4)*sum(k*(k+1)*exp(-epsilon*beta4*k),k=0..infinity);\\
B4:=simplify(A4);\\
C4:=series(B4,epsilon,6);\\
D4:=convert(C4,polynom);\\
F4:=subs(n4=n-l*n2+(1/2)*l$\widehat{\rule{3mm}{0mm}}$2*n2,D4);\\
G4:=subs(beta4=beta-l*beta1+(1/2)*l$\widehat{\rule{3mm}{0mm}}$2*beta2,F4);\\
H4:=subs(epsilon=l*gamma,G4);\\
J4:=subs(w4=w-l*w1+(1/2)*l$\widehat{\rule{3mm}{0mm}}$2*w2,H4);\\
K4:=series(J4,l,3);\\
L4:=convert(L4,polynom);\\
Z4:=simplify(L4);\\
ans1:=simplify(Z1+Z2+Z3+Z4);\\
ans2:=subs(w=V1/gamma,ans1);\\
ans3:=subs(w1=V2/gamma,ans2);\\
$\sharp$ Note that the result is of second degree in l and is independent\\
$\sharp$ of gamma.\\
ans4:=subs(beta=1/Theta(x),ans3);\\
ans5:=subs(beta1=diff(1/Theta(x),x),ans4);\\
ans6:=subs(beta2=diff(1/Theta(x),x\$2),ans5);\\
Ans:=simplify(ans6);\\
This gives us eq.~(\ref{eq-rhothetadot}).
\end{document}